# A Palmtop Synchrotron-like Radiation Source

**Authors:**

Min Chen[1,2,*], Fei-Yu Li[3], Ji Luo[1,2], Feng Liu[1,2], Zheng-Ming Sheng[1,2,3,†], & Jie Zhang[1,2]

**Affiliations:**

[1]Key Laboratory for Laser Plasmas (Ministry of Education) and Department of Physics and Astronomy, Shanghai Jiao Tong University, Shanghai, 200240, China

[2]IFSA Collaborative Innovation Center, Shanghai Jiao Tong University, Shanghai 200240, China

[3]SUPA, Department of Physics, University of Strathclyde, Glasgow G4 0NG, UK

[*]e-mail: minchen@sjtu.edu.cn

[†]e-mail: zhengming.sheng@strath.ac.uk

**Synchrotron radiation sources are immensely useful tools for scientific researches and many practical applications[1-4]. Currently, the state-of-the-art synchrotrons rely on conventional accelerators, where electrons are accelerated in a straight line and radiate in bending magnets or other insertion devices. However, these facilities are usually large and costly[5]. Here, we propose a compact all-optical synchrotron-like radiation source based on laser-plasma acceleration either in a straight or in a curved plasma channel. With the laser pulse off-axially injected in a straight channel, the centroid oscillation of the pulse causes a wiggler motion of the whole accelerating structure including the trapped electrons, leading to strong synchrotron-like radiations with tunable spectra. It is further shown that a ring-shaped synchrotron is possible in a curved plasma channel. Due to the intense acceleration and bending fields inside plasmas, the central part of the sources can be made within palm size. With its potential of high flexibility and tunability, such compact light sources once realized would find applications in wide areas and make up the shortage of large synchrotron radiation facilities.**

The laser plasma based new concept of accelerator, i.e. laser wakefield accelerator, has shown rapid progress in the last 30 years[6–8]. In this scheme, an ultrashort intense laser pulse is used to excite a large-amplitude plasma wave with field strength of about 100 GV/m and phase velocity close to the speed of light. Electrons can be trapped in such a structure and be accelerated to GeV energy in a centimeter distance[9, 10]. The new record of beam energy is 4.2 GeV, achieved recently by Leemans' group at Lawrence Berkeley National Laboratory in 9 cm acceleration distance by a 350 TW laser[11]. During the longitudinal acceleration, electrons in such a structure usually also undergo transverse betatron oscillations which lead to electromagnetic radiation[12, 13]. Betatron radiation has both been studied theoretically and demonstrated experimentally[14–17]. A peak brightness of $1\times10^{22}$ photons/s/mrad$^2$/mm$^2$/0.1%BW has been observed in experiments and simulations[18]. Another way to use such electron beam for radiation is by use of external insertion devices such as a magnetic wiggler or undulator[19–21], or a light undulator[22, 23]. Schlenvoigt *et al.* have reported their experimental results of 740nm radiation with a peak brilliance $6.5\times10^{16}$ photons/s/mrad$^2$/mm$^2$/0.1%BW, which is found from the interaction of a 28 pC, 10-fs-duration electron bunch with a 50 period, 1-m-long undulator with a deflection parameter *K*=0.6[20]. Although the electron betatron radiation inside a wakefield can generate high frequency radiation, the limited tunability in the light spectrum and brightness hinder its wide applications. The combination of LWFA with usual magnetic devices may produce high energy photons for applications, the size of the whole facility is, however, inevitably much expanded. The design of an all-optical synchrotron-like radiation device with improved properties would be extremely attractive for broad applications.

Here, we propose a new route towards synchrotron-like radiation with laser plasma based accelerators and undulators formed inside straight or curved plasma channels. Different from the normal betatron radiation of electrons oscillating in a wakefield, our scheme basically results from the laser centroid guiding and transverse wake structure oscillation. It gives controllable oscillation frequency, amplitude, electron beam and radiation beam pointing. These then provide more controllability for the radiation, important for practical applications. By using multidimensional particle-in-cell (PIC) simulations we show that both the driver laser pulse and the electron beam accelerated behind can be well guided over considerable distances inside the plasma channels. In the case with the straight plasma channel, off-axis or obliquely laser injection leads to transverse oscillations of the electron beams due to the centroid oscillation of the driver laser. In the case with the curved plasma channel, bent propagation of the driver laser and the trapped electron beams are also found. Synchrotron radiations are then generated naturally from the beam acceleration and guiding processes. Since the oscillating propagation and bent propagation of the beams as well as the beam energy inside the channels can be fully controlled by the laser and channel parameters, the radiation spectra can be largely tunable.

**Laser Wakefield Acceleration in a Straight Plasma Channel**
Due to the natural diffraction of a laser beam beyond the Rayleigh length ($Z_R = k_0 W_0^2/2$), a parabolic plasma channel with density profile $n(r) = n_0 + \Delta n r^2/r_0^2$ was suggested and used for long distance propagation of high power lasers[7]. Here, $k_0 = 2\pi/\lambda_0$ is the laser wave number, *W*$_0$ is the focal spot size, $\Delta n = n(r_0) - n_0$ is the channel depth and $r_0$ is its width. Plenty of theoretical studies have shown that a matched plasma channel can be used to keep laser spot size stable during the propagation. However, a perfect matching is almost impossible in reality. Usually the laser spot size experiences periodic oscillations during their propagation when the channel is not well matched with the laser spot size. This kind of mismatching can lead to wakefield evolution in LWFA, followed with electron injection. There are many works published on these effects.

Besides the laser spot size evolution, there is also another factor critical for LWFA in a plasma channel, which is rarely studied so far. It is about the laser injection position and direction relative to the plasma channel axis. If the laser is off-axis injected, i.e., the initial laser propagation direction is not along the channel axis, or the laser is obliquely incident into the plasma channel, the whole electron beam may oscillate transversely inside the plasma channel following the laser propagation. As we will show in this paper, the oscillation and guiding of the laser pulse in such cases can be used to induce wiggler/undulator like radiation from the electrons accelerated in the wakefield. We first study the propagation of an off-axis injection pulse, and then extend it to the oblique incidence case. Based upon these, a bent plasma channel is introduced in the next section. The radiation properties of electrons inside the channels will be discussed finally.

For a straight plasma channel, from the theory developed in Ref. [7], the spot size $r_s$ of a Gaussian laser beam with $|a|^2 = (a_0 r_0/g_s)^2 \exp(-2r^2/r_s^2)$ evolves according to

$$\frac{d^2 R}{dz^2} = \frac{1}{Z_M^2 R^3}\left(1 - \frac{\Delta n}{\Delta n_c} R^4\right), \tag{1}$$

where $R = r_s/r_0$ is the normalized spot size, $Z_M = \pi r_0^2/\lambda_0$, and $\Delta n_c = (\pi r_e r_0^2)^{-1} = 1.13 \times 10^{20}(cm^{-3})/r_0^2(\mu m)$ with $r_e = e^2/m_e c^2$ is the classical electron radius. For a matched plasma channel and laser pulse ($\Delta n = \Delta n_c$ and $r_s = r_0$), the laser spot size is stable during the propagation. Otherwise the spot size oscillates between $r_s = r_i$ and $r_s = (\Delta n/\Delta n_c)^{1/2} r_0^2/r_i$, where $r_i$ is the laser spot size at the entrance ($z = 0$) and $dr_s/dz = 0$ at $z = 0$ is assumed. The oscillation period length is $\lambda_{os} = \pi Z_M (\Delta n/\Delta n_c)^{1/2}$ (see Fig. 1(a)).

The above analytical results are valid for the co-axis injection case. If the laser pulse is off-axis injected, the laser centroid will also oscillate transversely besides the normal self-focusing and defocusing (see Fig. 1(b)). The oscillation period can be analyzed by comparing Figs. 1(a) and 1(b). If we trace the trajectory of the top half part of a laser pulse which initially injects on axis, after $2\lambda_{os}$ distance the top half part goes back to its initial transverse position. So it is reasonable to think the transverse oscillation period for an off-axis injection pulse is $\Lambda_{os} = 2\lambda_{os}$. We have carried out two-dimensional PIC simulations to confirm this.

In the 2D simulations the laser pulse injects into the simulation box from the left boundary. The normalized intensity is $a = eEm\omega_0 c = a_0 \times exp(-t^2/L_0^2 - r^2/W_0^2)$ with $a_0 = 2.0, L_0 = 9.48T_0, W_0 = 10.0\lambda_0$ and $\lambda_0 = 0.8\mu m, T_0 = 2\pi/\omega_0 \simeq 2.67fs$. The laser is s-polarized with electric field perpendicular to the simulation plane. The plasma density has a channel profile with $n_0 = 0.001n_c$, and $\Delta n = \Delta n_c$, where $n_c \simeq 1.7 \times 10^{21}/cm^3$ is the critical plasma density for the laser pulse. We have used ionization injection for controlled electron injection by setting a mixed preformed plasma composed of fully ionized He plasma (composing the plasma density of $n_e$) and partially ionized Nitrogen plasma with density of $n_{N^{5+}} = 5.0 \times 10^{-4} n_c$ [24–27]. The nitrogen is located from $x = 30\lambda_0$ to $x = 50\lambda_0$ with an upramp-plateau-downramp ($5\lambda_0 - 10\lambda_0 - 5\lambda_0$) profile. This kind of distribution is used to obtain quasi-monoenergetic electron acceleration which is widely adopted in recent ionization injection experiments and can be replaced by other controlled injection schemes.

In Fig. 1(c) we show the centroid oscillations of the laser pulse (thick solid line) and accelerated electron beam (thin dash-dotted line) along the longitudinal acceleration distance for different channel width $r_0$. In the simulations we set $\Delta n = \Delta n_c$ and the initial off-axis displacement $Y = 3\lambda_0$ when the channel width $r_0$ is varied. From the analytical results[7], the oscillation period for focusing and defocusing is $\Lambda_{os} = 1.58$ mm for our simulation parameters of $r_0 = W_0$. The simulation result shows $\Lambda_{os} \simeq 1.56$ mm which is close to the analytical result. This is just what we have deduced before (see Figs. 1(a) and 1(b)). For different channel widths, analytical results show $\Lambda_{os} \propto Z_M\sqrt{\Delta n_c/\Delta n} \propto r_0^2$. The simulations give similar results: $\Lambda_{os}(r_0 = 1.2):\Lambda_{os}(r_0 = 1.0):\Lambda_{os}(r_0 = 0.8) = 2.25:1.56:1.05$, which is close to $1.2^2:1.0^2:0.8^2$. All these simulation results show that the laser centroid oscillation can be well described by the channel guiding theory, which provides for controllable beam propagation.

Besides the laser beam oscillation, Fig. 1(c) also shows the oscillation of accelerated electron beam. Normally the electron beams oscillate along with the laser beams almost synchronously except for a slight delay. The electron beam deviates more from the channel axis than the laser beam and its oscillation amplitude is also larger than that of the laser beam. When the channel width reduces and the electron energy increases, this kind of synchronism is broken as shown by the green solid line and the violet dash-dotted line for the laser centroid and the electron beam, respectively. Figures 1(d) and 1(e) show snapshots of typical acceleration structure, electron beam and laser beam at different time. As one can see, the laser beam, acceleration bubble and accelerated electrons oscillate obviously inside the channel. This kind of beam oscillation is different from the one found by Popp *et al.*[28], where both the acceleration structure and electron beam also deviate from straight line due to laser front tilting. Moreover, the electron beam cannot go back to the initial transverse position in their case. In the simulation we also find the laser pulse may tear to be broken transversely and longitudinally due to stronger oscillation if the channel depth or the initial laser off-axis distance is too large.

Besides plasma channel properties, both the transverse injection deviation and laser angular pointing are critical parameters for laser beam propagation. We study these parameter effects by fixing the channel parameters and changing the initial laser off-axis injection position and injection angle. When discussing the off-axis effects, the laser injection angle is fixed to be $\theta = 0°$ or parallel to the channel axis. In Fig. 2(a) three different initial off-axis positions are studied. As we can see the oscillation frequencies of the laser centroid are the same for all these cases. They are the same as the analytical results discussed before. In addition, a typical evolution of the laser maximum intensity for the transverse off-axis of $Y = 3\lambda_0$ is shown by the green dashed line. Its oscillation period is half of the laser centroid oscillation as expected according to Figs. 1(a) and 1(b). The effect of injection pointing angle is shown in Fig. 2(b), where two simulation cases with injection angles $\theta = 1°$ and $\theta = 2°$ are given. The larger the injection angle, the larger oscillation amplitude of the laser beam. However, the oscillation period is independent of this parameter. It is only determined by the channel properties as shown in the analytical part. There is also some up-limit for the injection angle, beyond which the laser pulse may experience deformation and no suitable acceleration structure can be formed. For the given channel parameters in this work, our simulation shows that the laser pulse will go through the channel boundary transversely when $\theta \geq 5°$, and electrons are no longer accelerated along the plasma channel. In principle, the effect of injection angle is the same as the off-axis injection since the latter also leads to laser pulse propagation at some tilt angles inside the channel.

In reality, laser propagation is a kind of 3D phenomenon, especially the self-focusing is dimensional dependent. We have carried out 3D simulations for the off-axis injection of the laser pulse and corresponding electron beams in a plasma channel. Due to the limit of computational resources, we scale down the laser plasma parameters to a small acceleration structure by increasing the plasma density to $n_0 = 0.01n_c$ and correspondingly reduce the size of the laser pulse. The laser is initially off-axis injected into the plasma channel with the incident angle of zero degree. Figure 3 shows typical snapshots of the wakefield distribution, accelerated electron beam and longitudinal electric fields at two different acceleration distances. From the projections of the wakefield and accelerated electron beam, one can see both the laser pulse and electron beam oscillate in the plane determined by the initial laser propagation direction and the channel axis in the 3D case. In the meanwhile, 2D simulations for such parameters are also carried out, which shows similar results. This benchmarks the rationality of our simulations of using 2D geometry for other cases.

In addition, there is another universal case we have not included here for 3D simulation. The centroid of a laser pulse may perform spiral motion if the initial laser propagation axis is in an arbitrary skew angle against the channel axis. This is a natural deduction. The radiation by electrons from this motion may show unique characters. To avoid excessive expansion of our current paper, we leave these in future studies.

**Laser Wakefield Acceleration in a Curved Plasma Channel**
From the above analytical and simulation results of laser and beam centroid oscillations, it appears possible to guide both the laser pulse and electron beam even along a continuous bending trajectory by using an initially bent plasma channel with certain curvature. Once realized, it can lead to controllable pointing of an electron beam and the associated radiation, which may benefit the applications of LWFA accelerated electron beam and radiation. Even a closed or open ring structure may also be possible provided the laser power is high enough and the curvature of the bent plasma channel is proper, as schematically shown in Fig. 4(a). However, it is almost impossible to simulate laser guiding in a closed ring structure with the current computational ability and routine PIC codes, where the curvature of the ring is expected to be around centimeter scale or larger.

Even though one cannot simulate the whole laser guiding and electron acceleration processes with a curved plasma channel in a large range region, it is still possible to test this idea in a part of a ring structure (see Methods). The plasma parameters of the arc-shaped channel are chosen according to the simulation results given above for the case of a straight plasma channel. We set the outer and inner radiuses of the ring shape plasma channel boundary to be $R_1 = 30.04mm$ and $R_2 = 29.96mm$, respectively. The channel density parameters and laser parameters are the same as the 2D simulations as before. The total simulation regime covers a region of $9\text{mm} \times 1\text{mm}$. The bent plasma channel ends until $x = 7.76mm$, after which electrons and the laser pulse propagate inside vacuum for a short distance. To avoid electron losing due to continuous bending of the wake trajectory, a $0.76mm$ long straight pre-acceleration part is added before the channel goes into the bent region. In the pre-acceleration stage electrons can be accelerated to a small distance away from the end of the bucket so there will be enough transverse potential to trap the electrons avoiding transverse losing.

Typical simulation results are shown in Fig. 4 (b,c), where the evolution of the laser beam centroid and the energy of beam electrons and their average deflection angle are shown. As one can see, the trajectory is generally guided by the plasma channel. From the inset of Fig. 4 (b) we see the guided electron trajectory is beyond the channel center. In another simulation we find both the laser and electron beam trajectory may oscillate around the channel center as shown schematically in Fig. 4 (a) by the red dashed line. Figure 4(c) shows the electrons get continuous acceleration to 600MeV until the plasma channel ends at $x = 7.76mm.$ The moving angle of the electron beam is shown by the blue line, which also shows oscillating character. The final beam propagation direction is more than $12^o$ away from the initial injection direction. Two typical snapshots of acceleration structure at *x* = 0.632mm and *x* = 7.667mm are shown on top of Fig. 4(b). From them one can see the bending of the acceleration trajectory clearly. Longer acceleration along the ring structure is possible if one uses a high energy laser pulse along with an appropriate guiding channel. Electron acceleration at 10GeV level by using 10cm plasma channel is well studied theoretically, e.g., in LBNL. Further acceleration length is also under discussion in many laboratories around the world. To effectively bend an electron beam, the plasma channel length can be reduced by using a smaller curvature radius. However, it should be carefully selected otherwise the laser pulse may go through the channel boundary or be deformed seriously. The minimum value depends on channel depth and laser intensity. In our simulation parameters *R* = 2.4cm is still possible for laser and electron beam guiding, which means the circumstance is about 15cm, which is already close to the BELLA like plasma channel length. In addition, since electron energy is continuously increased until dephasing happens inside the curved plasma channel, a spiral like plasma channel with radius increasing with the polar angle may be needed to accomplish cyclotron like radiation.

**Controlled synchrotron radiation in straight and curved plasma channels**
From the above studies, one can see that both in straight or curved plasma channels electrons can experience controlled transverse acceleration. Similar to the electron beams transverse acceleration in insertion devices or bending magnets in a traditional storage ring, synchrotron-like radiations can be generated in such plasma channels. For incoherent radiations, the radiation intensity is proportional to the final electron beam charge. It depends on the injection process and beam loading effect during the wake acceleration. One may increase the final electron charge by increasing the concentration of Nitrogen when ionization injection is used. In our typical 2D simulations, $6 \times 10^6/\mu m \sim 3 \times 10^7/\mu m$ electrons are accelerated, which corresponds to about a few pC electrons. Besides the electron charge, the beam propagation trajectory is another key factor that affects the final radiation spectrum.

In a straight plasma channel, electron path can be approximated by a sinusoidal curve with a period of

$\Lambda_{os} = 2\pi^2 r_0^2 (\Delta n/\Delta n_c)^{1/2}/\lambda_0$ , which can be tuned from hundreds of micrometers to few centimeters long. The transverse oscillation amplitude depends on the initial laser off-axis distance ($Y$) which can be usually tuned up to a few micrometers. The energy of electron beam is continuously increased during the acceleration process until reaching the dephasing point. In our typical simulations electron energy can reach 450 MeV after 4.5mm acceleration distance. Correspondingly the oscillation strength parameter $K = 2\pi\gamma Y/\Lambda_{os}$ is widely tunable (around 0.1 to 10), which means the radiation can show both undulator (when $K$ < 1) or wiggler (when $K$ > 1) like spectrum. Typical radiation photon energy is $E_p = 2\gamma^2 hc/(1 + K^2/2 + \gamma^2\theta^2)\Lambda_{os}$, where $h$ is planck constant and $\theta$ is radiation angle.

Typical angular integrated radiation spectra (dI/dω) from the electrons are shown in Fig. 5(d). To see the tunability of the radiation, three cases are studied: (i) the plasma channel width $r_0=w_0$, laser off-axis distance $Y=3\lambda_0$ (corresponding to violet and red lines in Fig. 5(d)); (ii) the plasma channel width $r_0=w_0$, laser off-axis distance $Y=2\lambda_0$ (corresponding to blue and black lines in Fig. 5(d)); (iii) the plasma channel width $r_0=1.2w_0$, laser off-axis distance $Y=3\lambda_0$ (corresponding to green and light blue lines in Fig. 5(d)). The radiation spectra are calculated with the VDSR code[31] (see Methods) using traced electrons in PIC simulations. Trajectories for some of the selected electrons of case (i) and case (iii) are shown in Fig. 5(a) and (b), respectively. As one can see electrons show irregular oscillations overlapping on regular oscillations at the beginning of the acceleration (x<1.588mm). To avoid the uncertainty of trajectory selection at the beginning the radiation calculation position begins at x=1.588mm. Low frequency radiations before this position have been omitted. Radiation spectra after two different acceleration distances are shown. The green, violet and blue lines correspond to an acceleration distance of $L_{acce}$=1.5mm; light blue, red and black lines correspond to the acceleration distance of $L_{acce}$=2.25mm. As one can see the radiation spectra show wide bandwidth. The peak positions of spectra for the long acceleration show obvious blue shift due to higher electron energy along the acceleration (see the color variation along each trajectory). Comparing cases (i) and (iii), one can see that larger channel width introduces larger oscillation period of the electron beam ($\Lambda_{os}$) and smaller photon energy. For our simulation parameters, the peak radiation frequency locates between 0.895keV to 3.88keV. The high energy tail extends to tens of keV, which covers the soft x-ray to hard x-ray regime. Fig. 5(e) and (f) show angular resolved spectra for case (i) and (iii), respectively. As one can see that besides frequency shifting, radiation angular distribution also varies significantly with the laser plasma parameters. This provides the flexibility for radiation pointing.

In a bent plasma channel, the curvature of the electron trajectories depends on the curvature of the plasma channel which has a lower limitation determined by the laser guiding. Usually the curvature radius $R$ is in few centimeter scale. Different from the normal radiation in a storage ring, in our case the electrons are accelerated continuously when they radiate. A multiple period motion inside a ring like plasma structure may not be helpful or necessary for such a radiation source since the laser depletion and electron dephasing may happen before reaching the multiple period motion. The radiation spectrum is more like the continuous radiation spectrum from a single bending magnet without harmonic characters. The critical radiation frequency is about $\omega_c \sim \gamma^3 c/R$. For the case in Fig. 4, there are $\gamma \simeq 1200, R = 3cm$ at the end of the plasma channel. This corresponds to a critical photon energy of 7 keV.

## Summary

In summary, a scheme of compact synchrotron radiation is proposed based upon laser wakefield acceleration of electrons by use of different plasma channels. For a straight plasma channel, electron beam oscillation and subsequent radiation are realized by use of off-axis injection of laser pulses and obliquely injected laser pulses. On this basis, we show that bent laser propagation and electron acceleration inside a curved plasma channel are possible, which lead to a new type of compact synchrotron radiation. In our scheme both transverse and longitudinal acceleration comes from the laser wakefield, without any external acceleration fields and bending magnets. Note that the two kinds of beam lines suggested with straight plasma channels and curved plasma channels are similar to conventional linear undulator radiation and storage ring radiation. The acceleration and radiation parts of this kind of devices can be made within palm size and thereby be constructed with much reduced cost. Although the quality of such light sources may not be comparable immediately with existing SRs, it shows unique advantages for wide applications. Even though the current paper has demonstrated the principle concept of the new type of synchrotron radiations, it is obvious that there are still high potential and need to improve the quality of such light sources with detailed design for laser wakefield acceleration and channel guiding.

It deserves to point out that our channel guiding studies for both laser and electron beams can also benefit other radiation schemes. By using a plasma channel, all optical Thomson or Compton scattering may be realized more easily[29]. Two lasers can be incident into a plasma channel from two ends of the channel separately. An intense short pulse drives a wakefield accelerating electrons behind. The other low intensity long laser pulse oppositely propagates and interacts with the accelerated electron beam and makes Thomson scattering. By using this method, the difficulty of laser beam overlapping in space and time could be significantly reduced since both of them can be automatically guided by the plasma channel. Detailed studies on this topic still need to be carried out both numerically and experimentally. The curved channel guiding may also be helpful to reduce the difficulty of staged wakefield acceleration since lasers with different propagation directions can be used for different acceleration stages, which provides more flexibility for experimental arrangements.

## Methods

The standard particle-in-cell simulations are used to study the wakefield acceleration process. Both VLPL[30] and OSIRIS[31] PIC codes are used to perform the simulations. The simulations by these two codes are benchmarked with each other, which guarantee the correctness of simulation results. Both of the codes are fully relativistic electromagnetic code. For the curved plasma channel simulation, to save the transverse size of the simulation box, periodic boundary conditions in transverse directions are used both for particles and fields. The simulation box size is $80\mu m \times 192\mu m$, which is much smaller than the total simulation space ($9mm \times 1mm$) both in longitudinal and transverse directions. This is done by adjusting the plasma channel distribution according to the simulation box size and the transverse periodic conditions. To save computational time, the background He plasma is represented by electrons and protons only. The initial state for Nitrogen is $N_{5+}$ since ionization injection in our laser plasma parameters comes from $N_{5+}$ and $N_{6+}$ ions. We use 16 macro particles per cell for Nitrogen ions and 9 for electrons. The cell size is $0.032\lambda_0 \times 0.25\lambda_0$ and temporal resolution is $dt = 0.0315T_0$ for the 2D simulations. Numerical convergence is checked by increasing the macro particle number and the simulation resolution. The laser pulse is focused at $x = 20\lambda_0$ and incidents from the left boundary with initial center at $x = -30\lambda_0$.

The radiation spectra are calculated by the post process code VDSR[32]. We traced 100 electrons from the high energy bunch in PIC simulations and input the trajectories into the VDSR code. The time step for the trajectory output is $dt = 0.09375T_0$. The electrons are recorded once they have been ionized. The code then calculates each electron's radiation along the trajectories. The final spectrum is incoherent addition of the radiation from the 100 electrons. In the code by selecting different end positions along the trajectories, radiations from different acceleration lengths are calculated. The radiation is only calculated for simulations with p-polarized laser pulses. In 2D PIC simulation although electrons show similar trajectories in the simulation plane (comparing Figs. 5(a) and 5(c)) when p and s polarized laser pulses are used, electrons will have nonzero momentum along the third direction but without displacement along this direction when s polarized laser pulse is used, which may introduce inconsistence to the radiation calculation.

**References**
1. Huxley, H.E. *et al.* The use of synchrotron radiation in time-resolved X-ray diffraction studies of myosin layer-line reflections during muscle contraction. *Nature* 284 140-3 (1980).
2. Chen, L. Duerr, K.L. and Gouaux, E. X-ray structures of AMPA receptor Cone snail toxin complexes illuminate activation mechanism. *Science* 345 1021 (2014).
3. Lee, C.-H. *et al.* NMDA receptor structures reveal subunit arrangement and pore architecture. *Nature* 511 191 (2014).
4. Schoenlein, R.W. *et al.* Generation of femtosecond pulses of synchrotron radiation. *Science* 287 2237-2240 (2000).
5. Lightsources of the World. http://www.lightsources.org/regions
6. Tajima, T. and Dawson, J.M. Laser electron accelerator. *Phys. Rev. Lett.* 43, 267 (1979).
7. Esarey, E. Schroeder, C.B. and Leemans,W.P. Physics of Laser-driven plasma-based electron accelerators. *Rev. Mod. Phys.* 81, 1229-1285 (2009).
8. Hooker, S.M. Developents in laser-driven plasma accelerators. *Nature Photon.* 7, 775-782 (2013).
9. Leemans, W.P. *et al.* GeV electron beams from a centimetre-scale accelerator. *Nature Phys.* 2 696-699 (2006).
10. Wang, X. *et al.*, Quasi-monoenergetic laser-plasma acceleration of electrons to 2GeV. *Nature Commun.* 4:1988, DOI:10.1038/ncomms2988 (2013).
11. Leemans, W.P. Gonsalves, A.J. Mao, H.-S. *et al.* Multi-Gev Electron Beams from apillary-Discharge-Guided Subpetawatt Laser Pulses in the Self-Trapping Regime. *Phys. Rev. Lett.* 113, 245002 (2014).
12. Corde, S. Phuoc, K. Ta Lambert, *et al.* Femtosecond x rays from laser-plasma accelerators. *Rev. Mod. Phys.* 85, 1 (2013) and reference therein.
13. Esarey, E. Shadwick, B.A. Catravas, P. and Leemans,W.P. Synchrotron radiation from electron beams in plasma-focusing channels. *Phys.Rev. E* 65 056505 (2002).
14. Schnell, M. Saevert, A. Uschmann, I. *et al.* Optical control of hard X-ray polarization by electron injection in a laser wakefield accelerator. *Nature Commun.* 4:2421

doi:10.1038/ncomms3421 (2013).

15. Kiselev, S. Pukhov, A. and Kostyukov I. X-ray generation in strongly nonlinear plasma waves. *Phys. Rev. Lett.* 93 135004 (2004).
16. Rousse, A. *et al.* Production of a keV x-ray beam from synchrotron radiation in relativistic laser-plasma interaction. *Phys. Rev. Lett.* 93 135005 (2004).
17. Kneip, S. *et al.* Observation of Synchrotron Radiation from Electrons Accelerated in a Petawatt-Laser-Generated Plasma Cavity. *Phys. Rev. Lett.* 100 105006 (2008).
18. Kneip, S. *et al.* Bright spatially coherent synchrotron X-rays from a table-top source. *Nature Phys.* 6 980-983 (2010).
19. Nakajima, K. Towards a table-top free-electron laser. *Nature Phys.* 4, 92-93 (2008).
20. Schlenvoigt, H.-P. *et al.* A compact synchrotron radiation source driven by a laser-plasma wakefield accelerator. *Nature Phys.* 4 130-133 (2008).
21. Fuchs, M. *et al.* Laser-driven soft-X-ray undulator source. *Nature Phys.* 5 826-829 (2009).
22. Phuoc, K.Ta *et al.* All-optical Compton gamma-ray source. *Nature Photon.* 6 308-311 (2012).
23. Cipiccia, S. *et al.* Gamma-rays from harmonically resonant betatron oscillations in a plasma wake. *Nature Phys.* 7 867-871 (2011).
24. Chen, M. Sheng, Z.M. Ma, Y.Y. Zhang, J. Electron injection and trapping in a laser wakefield by field ionization to high-charge states of gases. *J. Appl. Phy.* 99, 056109,(2006).
25. Chen, M. Esarey, E. Schroeder, C.B. *et al.* Theory of ionization-induced trapping in laser-plasma accelerators. *Phys. Plasmas* 19, 033101 (2012).
26. Liu, J.S. *et al.* All-Optical Cascaded LaserWakefield Accelerator Using Ionization-Induced Injection. *Phys. Rev. Lett.* 107 035001 (2011).
27. Pak, A. *et al.* Injection and Trapping of Tunnel-Ionized Electrons into Laser-Produced Wakes. *Phys. Rev. Lett.* 104 025003 (2010).
28. Popp, A. *et al.* All-Optical Steering of Laser-Wakefield-Accelerated Electron Beams. *Phys. Rev. Lett.* 105 215001 (2010).
29. Chen, S. *et al.* MeV-Energy X Rays from Inverse Compton Scattering with Laser-Wakefield Acelerated Electrons. *Phys. Rev. Lett.* 110, 155003 (2013).
30. Pukhov, A. Three-dimensional electromagnetic relativistic particle-in-cell code VLPL (virtual laser plasma lab) *J. Plasma Phys.* 61425C433 (1999).
31. Fonseca, R.A. *et al.* OSIRIS: A Three-Dimensional, Fully Relativistic Particle in Cell Code for Modeling Plasma Based Accelerators. *Lect. Notes Comput. Sci.* 2331 342C351 (2002).
32. Chen, M. Esarey, E. Geddes, C.G.R. *et al.* Modeling classical and quantum radiation from laser-plasma accelerators. *Phys. Rev. Spec.Topics-Acce. Beams* 16, 030701 (2013).

**Acknowledgments**

This work is supported in part by the National Basic Research Program of China (Grants 2013CBA01504) and the National Science Foundation of China (Grant Nos. 11421064, 11374209, 11374210). M.C. appreciates supports from Shanghai Science and Technology Commission (Grant No. 13PJ1403600) and National 1000 Young Talent Program. The authors would like to acknowledge the OSIRIS Consortium, consisting of UCLA and IST (Lisbon, Portugal) for the use of OSIRIS and the visXD framework. Simulations were performed on the Π Supercomputer at Shanghai Jiao Tong University.

**Author contributions**

M.C. conceived the idea; M.C., F.Y.L., F.L., J.L. and Z.M.S. performed the PIC simulations and radiation calculations; M.C., Z.M.S., F.Y.L. and J.Z. analyzed data and wrote the paper. All authors commented on the manuscript and agreed on the contents.

**Additional information**

**Competing financial interests:** The authors declare no competing financial interests.

Figures:

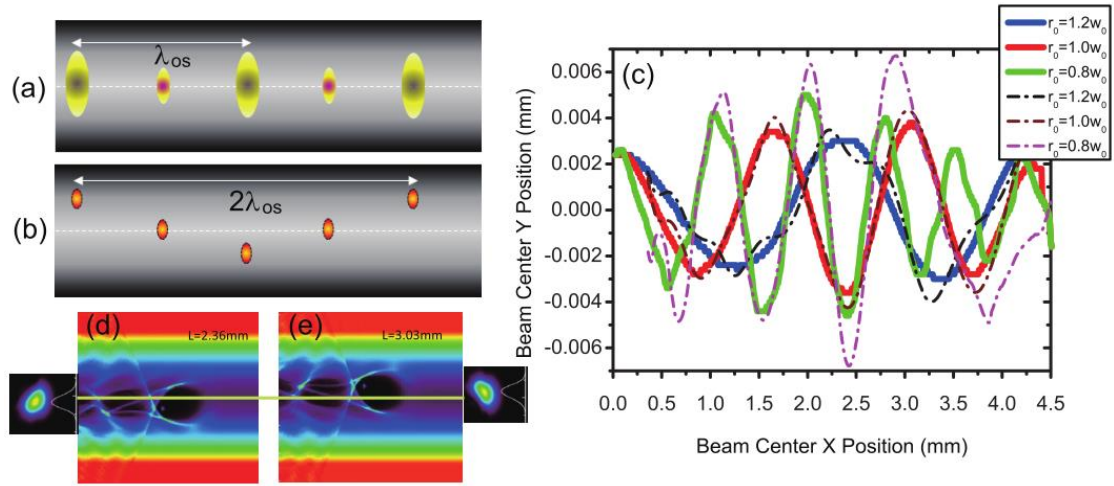

**Figure 1| Off-axis injection effects on laser wakefield acceleration in a straight plasma channel.**
(**a**) Schematic plot showing the laser propagation with on-axis injection in the plasma channel. (**b**) Laser propagation with off-axis injection in the plasma channel. The solid lines in (**c**) show the evolution of the laser central transverse position along their propagation distance under different channel radius $r_0$, while the dashed line show the central transverse position evolution of the accelerated electron beam along the longitudinal acceleration distance. (**d**) and (**e**) show snapshots of typical wakefield structures and accelerated beams at acceleration length of 2.36 mm and 3.03 mm, respectively, along with the corresponding laser beam profiles shown at two sides.

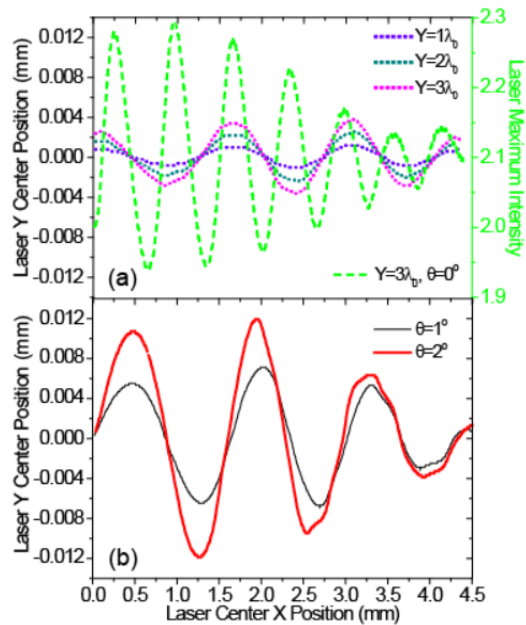

**Figure 2| Laser off-axis and oblique injection effects on beam oscillation.** (a) Change of the laser centroid along the propagation distance for off-axis injection with different initial deviation distances Y at the injection angle $\theta = 0°$. A typical laser intensity oscillation for the off-axis injection case of $Y = 3\lambda_0$ is shown by the green dashed line. (b) Change of the laser centroid along the propagation distance for oblique injection angles at $\theta = 1°$ and $\theta = 2°$ and with initial deviation distance Y=0. The laser and plasma density parameters are the same as before and $r_0 = W_0$ for all these cases.

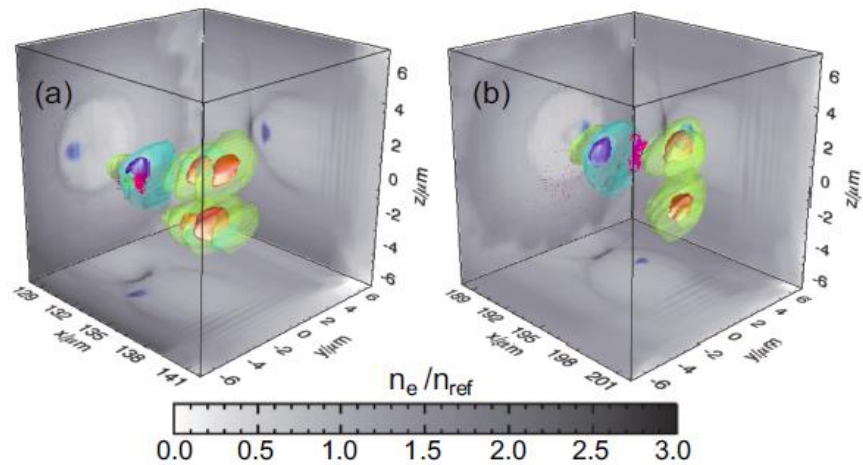

**Figure 3| Typical 3D-PIC simulation results for a wakefield acceleration from an off-axis injected laser beam.** Snapshots of plasma density (gray background), injected electron beam (red points) and longitudinal electric fields (blue-green-orange iso-surfaces) are shown for two different time steps. The projections on three surfaces of the cubic show the electrons oscillations along the acceleration distance. To save computational cost, laser plasma parameters are scaled down with $L_0 = 3T_0, W_0 = 4.5\lambda_0, Y = 3.0\lambda_0, \theta = 0.0°, n_0 = 0.01n_c$. The nitrogen density is $n_N = 5.0 \times 10^{-4} n_c$.

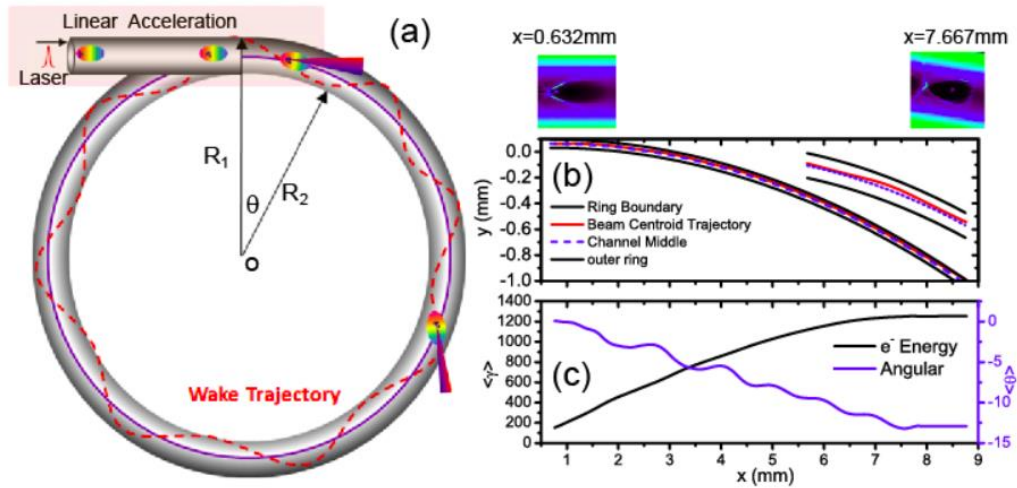

**Figure 4| Electron acceleration in a curved plasma channel.** (**a**) Schematic view of a synchrotron radiation ring based upon laser plasma wakefield acceleration, where the laser pulse propagates along the ring and oscillates around the channel axis with a typical trajectory shown by the red dashed line. (**b**) The trajectory of the laser pulse centroid through an arc-shaped channel in x-y plane from 2D PIC simulation. In the simulation the radius of the plasma channel is $R = (R_1 + R_2)/2 = 30 mm$. The top right inset of the figure shows the zoom in of a part of the trajectory. One can see the laser centroid actually is only slightly deviated from the plasma center in our simulation parameters. (**c**) The average electron energy and deflection angle are shown. On top of (**b**), two snapshots of the wakefields are shown, from which one can see the wake is deflected due to the curvature of the plasma channel.

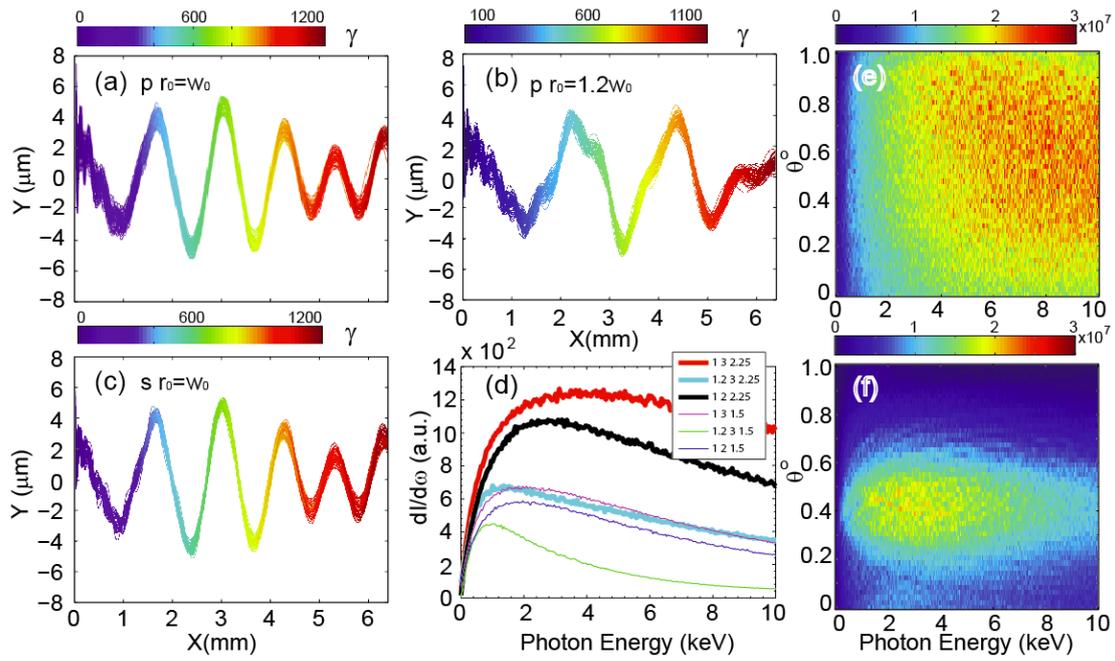

**Figure 5 | Typical trajectories of accelerated electrons and radiation spectra.** The colors represent the electrons' energies in (**a**), (**b**) and (**c**). (**a**) The laser is p-polarized and the channel width is $r_0=W_0$. (**b**) The channel width is $r_0=1.2W_0$. (**c**) The laser is s-polarized and the channel width is $r_0=W_0$. All the other simulation parameters are the same as those in Fig. 1. (**d**) Radiation spectra of the selected electrons. The radiation is calculated for a part of the trajectories. The upright legends represent the simulation parameters:[$r_0/W_0,Y/\lambda_0,L_{acce}$]. (**e,f**) Radiation distribution $dI^2/d\omega d\Omega$ (arbitrary unit) for case (i,iii), respectively. The acceleration distance here $L_{acce}$=3mm.